# Straintronics in Phosphorene: Tensile *vs* Shear Strains and Their Combinations for Manipulating the Band Gap

Anastasiia G. Solomenko[1], Ihor Y. Sahalianov[2], Taras M. Radchenko[1,*], and Valentyn A. Tatarenko[1]

[1] Department of Metallic State Theory, G. V. Kurdyumov Institute for Metal Physics of the N.A.S. of Ukraine, UA-03142 Kyiv, Ukraine
[2] Laboratory of Organic Electronics, Department of Science and Technology, Linköping University, SE-60174 Norrköping, Sweden
*tarad@imp.kiev.ua

## Abstract

We study the effects of the uniaxial tensile strain and shear deformation as well as their combinations on the electronic properties of single-layer black phosphorene. The evolutions of the strain-dependent band gap are obtained using the numerical calculations within the tight-binding (TB) model as well as the first-principles (DFT) simulations and compared with previous findings. The TB-model-based findings show that the band gap of the strain-free phosphorene agrees with the experimental value and linearly depends on both stretching and shearing: increases (decreases) as the stretching increases (decreases), whereas gradually decreases with increasing the shear. A linear dependence is less or more similar as compared to that obtained from the *ab initio* simulations for shear strain, however disagrees with a non-monotonic behaviour from the DFT-based calculations for tensile strain. Possible reasons for the discrepancy are discussed. In case of a combined deformation, when both strain types (tensile/compression + shear) are loaded simultaneously, their mutual influence extends the realizable band gap range: from zero up to the values respective to the wide-band-gap semiconductors. At a switched-on combined strain, the semiconductor–semimetal phase transition in the phosphorene is reachable at a weaker (strictly non-destructive) strain, which contributes to progress in fundamental and breakthroughs.

# Introduction: background of the problem and motivation of the study

The post-graphene era of two-dimensional or quasi-two-dimensional (2D) materials[1] got an additional impact for further development when in 2014 two different groups[2,3] independently exfoliated the single-layer black phosphorus — called "phosphorene" (further in the text mentioning phosphorene, we will mean black one) — from the bulk black phosphorus, which in turn has been first synthesized more than century ago[4,5]. Phosphorus is one of the abundant chemical elements in the Earth's crust (up to ≈0.1%)[6,7] and black P (α-form) is the most thermodynamically stable at ambient conditions among other phosphorus allotropes (white, red, violet, and A7 phase)[8,9]. Since 2014, extensive studies are carried out to promote the investigations of phosphorene: hundreds and even thousands of papers dealing with this material have been already published all over the world (*e.g.*, each of the Web of Science and Scopus scientometric databases numbers almost two thousands of articles containing the word "phosphorene" straight in the title).

As distinct from graphene, which is flatness (atomically flat), the crystal structure of phosphorene represents a corrugated atomic monolayer (see Fig. 1a–d), where chains of covalently bonded P atoms reside in two different planes. Among the family of currently known 2D materials, the monolayer black phosphorus attracts attention as a promising candidate not only for (opto)electronics, but for all material science as an interesting object for detailed study due to its peculiar features. Phosphorene has a natural direct band gap in the centre (Γ-point) of the Brillouin zone (Fig. 1e); however, its calculated value strongly differs in the literature from 0.76 to 2.31 eV (see collected data in tables [10,11] and Çakır *et al.*[12]) depending on the computational methods and approximations. At the same time, the experimentally measured gap value turned out also quite different: from 1.45 eV[2] to 2.05 eV[13] and 2.2 eV[14] (whereas for bulk black phosphorus, the gap is 0.31–0.36 eV[15–18]). Some authors claimed even higher ban gap values: up to 2.2 eV. Phosphorene exhibits a high on/off current ratio (up to ~$10^5$)[19] and (ambipolar)[20] carrier mobility (from 600 $cm^2V^{-1}s^{-1}$ at room temperature[21] up to ~$10^3$ $cm^2V^{-1}s^{-1}$ at 120 K and even higher at lower temperatures[22], *i.e.*, comparable with graphene). It is most remarkable and that is why worthy of attention features are high anisotropy (in mechanical, electronic, optical, thermal, and transport properties)[19,23–25], as a response to the anisotropy of the puckered (called also as buckled or wrinkled) lattice, and superior mechanical properties[26,27].

Due to the extraordinary mechanical flexibility, phosphorene can sustain the elastic deformations up to (its failure limits) ≈27–30% at single-axial strain[26] (which corresponds to the stress up to ≈10 N/m[27] depending on the stretching direction, and up to ≈10–13% at shear strains[28,29]. However, other authors claim about non-destructive ≈48% (≈11%) uniaxial deformation along armchair (zigzag) direction[24] and 30–35% shear strains[30]. If the isotropic strain reaches 29%, the puckered phosphorene structure transforms into a plane hexagonal flat lattice[31]. On one hand, such an enormous mechanical flexibility indicates the potential application at extreme conditions. On another hand, it offers an opportunity to consider this material (following graphene)[32,33] as an eminently suitable object of the study within the still relatively new research field in the condensed matter physics known as "straintronics"[34–37], in which the strain-induced physical effects in solids act as a tool for tuning (even amendment if any) the electronic properties. In the context of unique mechanical properties of phosphorene, another intriguing feature is its ability to be an auxetic, which is originated from its puckered structure. The existence of a negative Poisson's ratio $\nu$ in the out-of-plane direction under the uniaxial stretching in the direction parallel to the pucker, *i.e.*, along zigzag ($x$ as denoted in Fig. 1) direction, was revealed in several works[31,38–40], where a great difference of $\nu$ in different directions reflects the strong structural anisotropy. It is considered that phosphorene is a first known 2*D* auxetic material.

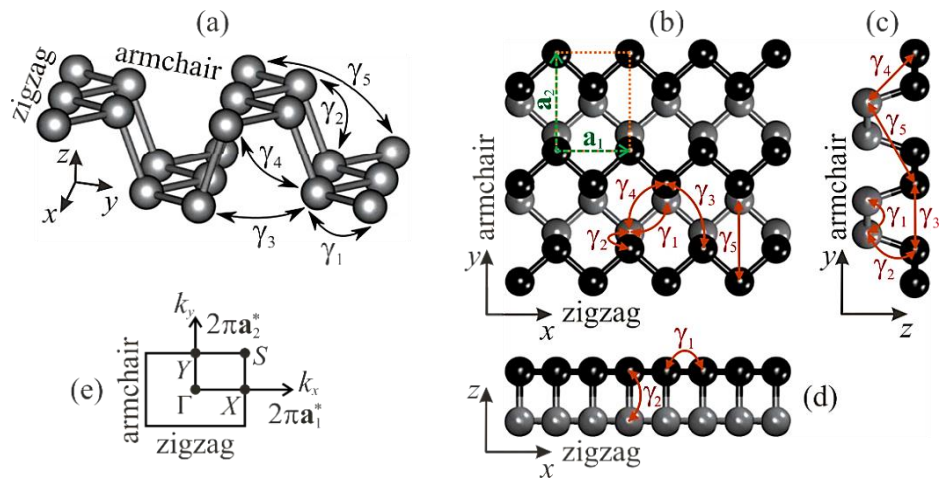

**Figure 1.** Crystal lattice of black phosphorene: perspective (a), top (b), and side (c, d) views. Black and grey balls in (b–d) depict the upper and lower sublayers of P atoms, respectively. Dashed rectangle in (b) outlines the basic unite cell (without taking into account the vacuum) with the basis translation vectors $\mathbf{a}_1$ and $\mathbf{a}_2$. The five non-equivalent hopping parameters in (a–d) are denoted by $\gamma_1$, $\gamma_2$, …, $\gamma_5$. Solid rectangle (e) exhibits the first Brillouin zone projected on *xy*-plane, where Γ, *X*, *Y*, *S* are the non-equivalent high-symmetry points in the reciprocal **k**-space with the basis translation vectors $\mathbf{a}_1^*$ and $\mathbf{a}_2^*$.

Similarly to its forefather, — graphene,[41,42] — the sensitivity of phosphorene's electronic structure to the external perturbations has noticed attention of researchers and as a result caused quite a number of mainly theoretical (analytical or/and computational) studies dealing with the response of the band gap behaviour on the mechanical field influence. Such a field is reflected in the axial (uni-, bi-, and even tri-axial) and isotropic tensile (or compression as a negative extension) strains. Rodin *et al.*[43] and Hien *et al.*[44] predicted the gap closing upon moderate compression along the transverse (out-of-plane) $z$ direction in phosphorene (see Fig. 1a–d). However, according to Li *et al.*[45], for the band gap closing, the biaxial compressive strain of 9% is required, while Hu *et al.*[31] assert about isotropic 7% compression or 22% stretching for the gap closing. Appearance of the semi-metallic and insulating (more than 3 eV gap) phases in phosphorene at the uniaxial + biaxial deformations in the range of ±20% along the armchair (Γ–Y) and zigzag (Γ–X) directions (see Fig. 1e) predicted by Yarmohammadi with co-authors[46], where they obtained the band gap for unstrained phosphorene as 1.52 eV. In contrast to them, some other authors[31,38,43,47] claimed different band gap values (at the Γ-point) of unstrained phosphorene: from 0.7–0.95 eV[31,39,43,47] up to 2.31 eV[12] depending on the approximations and computation schemes have been used. Phuc *et al.*[47] also observed a semiconductor–semimetal transition at the compression of 13% or 10% along the armchair and zigzag direction, respectively; moreover, even transition to the metallic state at the larger compressions. This disagrees with Elahi *et al.*[39], who reported about other strains of such a transition. In addition to the magnitude of the band gap, its type (direct–indirect) also undergoes the deformation-induced changing from direct to indirect and vice versa as many authors argue[27,31,39,47,48]. They report a whole number of cases when they observed such a transition for different strain types and values. Sometimes these values and types of such a critical strain even contradict each other. Interestingly, that appropriate interplanar (*e.g.*, uniaxial or biaxial) strain can even rotate the preferred electrical conducting direction in phosphorene[49].

There is still a lack of studies of phosphorene under elastic shear strain — another important and commonly occurred external stress for graphene[50,51] and other members of the 2*D* materials' family[52]. As well as the tensile deformation, such a strain type can be realized in different forms, *e.g.*, global shear deformation, process-induced or post-processing shearing[28]. At present, we found only three works (two computational[28,30] and one analytical[53]) reporting on the influence of shear strain on electronic structure and properties of phosphorene. The case of simultaneous effect of both (tensile and shear) strain types is found in one work[53] only, where authors reported that the most effective, *i.e.*, large, band gap manipulation does not require strongly armchair or zigzag direction. They[53] claim that the optimal strain direction depends rather on its type, and predict the combination of uniaxial armchair strain and shear deformation as the most effective approach to enlarging the bandgap. The statement about insignificance of the strain direction on its effectiveness[53] contradicts to the statement of Wang *et al.*[54] that the gap enlarges with the changing the direction of stress from zigzag to armchair.

Summarizing the results, we analysed in the above-mentioned briefly reviewed articles, one can emphasis on the following moments. (i) The lion's share of them reports on the results obtained from computations. In such computational studies, authors used commonly the VASP[2,10,27,28,45,55] or

Quantum ESPRESSO[30,43,47,49] and rarely other simulation packages like SIESTA[2,12,56], ABINIT[48] or VNL-ATK [57] for performing the density-functional theory first-principles calculations. Such calculations are suitable, accurate, and fruitful, however computationally expensive since require high computational capabilities (even implemented in the author's program code [58–60] ). *Ab initio* calculations are not feasible for large or moderate scale systems. Therefore, the phosphorene computational-domain sizes in such calculations are restricted to one or several unite cells, periodic supercells or lattice fragments with a relatively small (up to several dozens) quantity of the lattice sites (atoms). (ii) Although all authors report on the phosphorene as a superconductor with moderate energy gap, the reported gap values differ from each other more than 150% (even for the unstrained sample), which seems quite a lot. The calculated values depend on the applied computational methods and model approximations. All authors claim about deformational tunability of the band gap, however there is a lack of uniqueness on the strain type and values resulting in increasing (decreasing) the gap up (down) to the insulator (semimetal) state. (iii) There is a deficiency in the study of the shear strain effects on the phosphorene electronic structure, particularly on its band gap behaviour, especially as because this strain type can be dominated in the 2*D*-sheets-based flexible electronics[61].

Motivated by the restrictions, discrepancies, and scarcity indicated above in pts. (i)–(iii), respectively, in this work we use the analytical model of the tight-binding approximation with distant-dependent hoping integrals and Green's function method for implementation into the efficient numerical methodology of calculation of density of electronic states followed by the extraction of the band gap in phosphorene subjected to the uniaxial intraplanar tensile and shear strains as well as their combinations. At that, our computational domain contains millions of atoms, *i.e.*, close to realistic samples. In addition, we also use the first-principles calculations based on a density functional theory implemented in the Quantum Espresso package to compare our results obtained within both methods as well as with findings of other authors.

## Theoretical and numerical approaches

### Tight-binding Hamiltonian

Within the framework of the effective tight-binding (TB) model, the Hamiltonian in a real-space representation is defined on a phosphorene lattice as[44,46,62–65]

$$\hat{H} = \sum_{i \neq j} \gamma_{ij} c_i^\dagger c_j + \text{H.c.} = \gamma_1 \sum_{i \neq j} c_i^\dagger c_j + \gamma_2 \sum_{i \neq j} c_i^\dagger c_j + \gamma_3 \sum_{i \neq j} c_i^\dagger c_j + \gamma_4 \sum_{i \neq j} c_i^\dagger c_j + \gamma_5 \sum_{i \neq j} c_i^\dagger c_j + ..., \quad (1)$$

where indices *i* and *j* run over all phosphorene-lattice-sites, $c_i^\dagger$ (*c_j*) is the fermionic creation (annihilation) operator of electrons at *i* (*j*) site, and $\gamma_{ij} = \gamma_{ji}$ are the hopping parameters between the sites *i* (*j*) and *j* (*i*), and H.c. reflects the Hermitian conjugate of creation and annihilation operators. In our model, we take into account five nearest neighbour hopping integrals $\gamma_1$, $\gamma_2$, $\gamma_3$, $\gamma_4$, $\gamma_5$ denoted in Fig. 1a–d; thus, the summation in Eq. (1) is carried out up to the fifth neighbours. Note that the Hamiltonian in Eq. (1) does not contain any on-site term, in contrast to two papers by Katsnelson with co-authors[66,67], since here, we consider the defect-free lattice, so electrons have equivalent energies over all sites.

The values of the tight-binding parameters for the unstrained phosphorene lattice were adopted from Rudenko *et al.*[66]: $\gamma_1^0 = -1.22$ eV ($l_1^0 = 0.222$ nm), $\gamma_2^0 = 3.665$ eV ($l_2^0 = 0.224$ nm), $\gamma_3^0 = -0.205$ eV ($l_3^0 = 0.334$ nm), $\gamma_4^0 = -0.105$ eV ($l_4^0 = 0.347$ nm), $\gamma_5^0 = -0.055$ eV ($l_5^0 = 0.423$ nm), where superscript indicates that the system is unstrained (*i.e.*, relaxed or optimized), and values in the brackets—corresponding distances between the lattice sites (see Fig. 1a–d). Authors[66] used so-called *GW* approximation to obtain the parametrized *GW* Hamiltonian matrix and then extracted hopping energies from this matrix represented in the Wannier-functions' basis $\langle w_i | \hat{H} | w_j \rangle$. Latter these values were also adapted in quite a few other studies[44,46,64,65,68]. Midtvedt *et al.*[63] used a valence-force field model[69] to obtain an alternative set of the hopping parameters applicable for small and moderate strains <5%.

### Strain-caused hopping replacement

We treat of two types of the homogeneous elastic intraplanar deformations: single-axial (Fig, 2a, b)

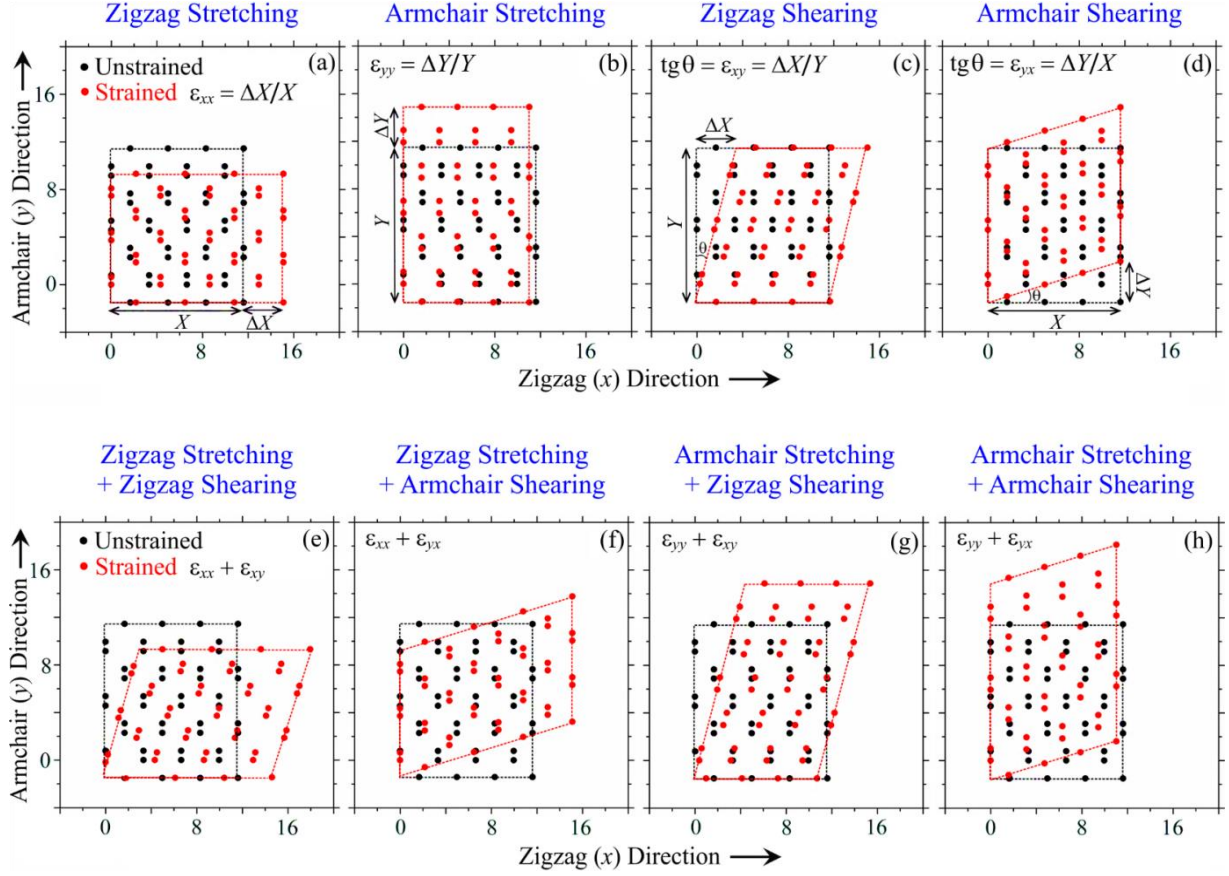

**Figure 2.** (Colour online) Schematic top views of unstrained and strained black phosphorene lattice as a result of the uniaxial tensile (a, b) or shear (c, d) deformations and their combinations (e–h).

and shear (Fig. 2c, d) strains. For both types, two orthogonally related directions of the applied strain are examined: along the zigzag (Fig. 2a, c) and armchair (Fig. 2b, d) edge. Since in a realistic case, the deformed lattice may be a result of a combination of several strain types, the combined deformation types, — stretching + shearing along one of or both directions (Fig. 2e–h), — are also included.

The uniaxial or/and shear strains induces a lattice deformation for both mutually transverse directions (as well as for any other one), *i.e.*, changes in the interatomic distances (bond lengths $l$) and bond angles. These values, bond lengths and bond angles, can be changed slightly or strongly depending on the strain type, magnitude and direction; nevertheless, they take the consequences of the change in the hopping integrals between the lattice sites. In a general case, the hoppings can differ among different neighbouring sites. Since we consider a uniform elastic strain, different hopping integrals from a given site to its neighbours should be the same for every such a site. Thus, the model Hamiltonian in Eq. (1) includes five distinct hoppings. As for to the strain-affected graphene[34,35,41], we follow the strain-displacement relations[63] for a homogeneous elastic strain in phosphorene that can be obtained from a valence-force model[70] and assume an exponential dependence of the hopping parameters on the interatomic P–P distances $l$:[63]

$$\gamma_{ij}(l) = \gamma_{ij}^0 \exp\left(-\beta\left(|\mathbf{R}'_{ij}|/|\mathbf{R}_{ij}|-1\right)\right), \quad (2)$$

where $\gamma_{ij}^0$ is a hopping of unstrained lattice, $\mathbf{R}_{ij}$ ($\mathbf{R}'_{ij}$) is an original (modified) vector connecting atoms at $i$- and $j$-site, $\beta \approx 1.2$ quantifies the decay rate[71], and a transverse response to the applied uniaxial strain (Poisson's effect) is also included. (An alternative way[44,46,53,64,68,72] to include the strain effect on the hopping parameters in a linear deformation regime is based on the Harrison's rule[73] defining the inversely quadratic dependence of the hopping energies on the bond length: $\gamma(l) \propto 1/l^2$.)

Table 1 collects data from several works[26,27,38,39], where authors reported on the calculated Poisson's ratios. In contrast to graphene, for which the Poisson's ratio $\nu$ is isotropic, in case of

**Table 1.** In-plane $v_x(y)$, $v_y(x)$ and out-of-plane $v_z(y)$, $v_z(x)$ Poisson's ratios as the responses on the uniaxial stretches oriented along armchair (*y*) and zigzag (*x*) directions, respectively, in phosphorene.

|  | Uniaxial strain direction | | Reference |
|---|---|---|---|
|  | Armchair (*y*) | Zigzag (*x*) |  |
| In-plane ν: $v_x(y)$ and $v_y(x)$ | 0.17<br>0.20<br>0.40<br>0.24 | 0.62<br>0.70<br>0.93<br>0.81 | Wei et al.[26]<br>Peng et al.[27]<br>Jiang et al.[38]<br>Elahi et al.[39] |
| Out-of-plain ν: $v_z(y)$ and $v_z(x)$ | 0.82*<br>0.046<br>0.210 | 0.50*<br>−0.027<br>−0.090 | Wei et al.[26]<br>Jiang et al.[38]<br>Elahi et al.[39] |

* The out-of-plain Poisson's ratios referred to Wei et al.[26] are calculated from the ratio of distances between the puckered upper and lower sublayers of the relaxed (*i.e.*, unstrained) phosphorene and uniaxially strained by 30%.

phosphorene ν is different for two in-plane (in-layer) stretching directions (along zigzag, *x*, and armchair, *y*): $v_y \neq v_x$. Moreover, the out-of-plane (out-of-layer) Poisson's ratio $v_z \neq v_y \neq v_x$ appears and depends on the in-layer uniaxial tensile direction as well. Despite an insignificant inconsistency, the values in Table 1 adequately exhibit an anisotropy in a mechanical response of phosphorene, and a negative out-of-layer Poisson's ration is revealed[38,39] along the armchair (*y*) direction in response to perpendicular (zigzag, *x*) uniaxial stretch (at least in the strain range of $-5\% \leq \varepsilon_{xx} \leq 5\%$)[38]. In our calculations, we use the in-plain ν averaged over those presented in Table 1 for every strain orientation, and the out-of-plain ν averaged over data of Jiang et al.[38] and Elahi et al.[39] for small or moderate strains ($\leq \pm 5\%$) and adopted from Wei et al.[26] for larger deformations (up to the predicted failure limit).

### TB-model-based computational details

The density of states (DOS) ρ for each energy level *E* per unit area *S* and spin relates is defined as

$$\rho(E) = \text{Tr}\left(\delta(E - \hat{H})/S\right), \quad (3)$$

where $\hat{H}$ is the Hamiltonian matrix given in Eq. (1). To calculate the DOS of the system, we implement an efficient numerical methodology or algorithm to be more precise, which has been developed earlier for graphene[41,74].

The total DOS ρ(*E*) can be represented as a sum of local density of states (LDOS): $\rho(E) = \sum_i^N \rho_i(E)$ with summation over all phosphorene lattice sites *N*. The LDOS relates to the imaginary part of the diagonal elements of Green's function[44,75,76] as $\rho_i(E) = -\pi^{-1} \text{Im} G_{ii}(E + i\zeta)$, where small ζ coefficient smooths peaks in the DOS separated by the energy ranges $\Delta E \propto 1/N$. To calculate the diagonal elements $G_{ii}$, we have to perform the tridiagonalization procedure for the Hamiltonian (1), which requires the most computational time and depends on the number of sites *N*. To calculate the first diagonal element of the Green's function $G_{11}$, we use the continued fraction technique. In principle, if we calculated LDOS at the first site, $\rho_1(E)$, we can repeat such calculations for all other (remaining) *N* − 1 sites in order to obtain the total DOS; however, this procedure requires the computational time quadratically dependent on the size of the system ($\propto N^2$) at issue. Thus, for the realistically large systems with millions of sites (atoms), another method in which the total computational efforts remain to be of the order of *N* is more reasonable and efficient. Such a method consists in an idea that a sufficiently large subsystem of the total system possesses the same DOS as an original (total) system. We choose a subsystem of Δ*N* sites at a phosphorene lattice and construct a wave function (packet) $|\psi_{rnd}\rangle \equiv \Delta N^{-1/2} \sum_i \exp(2\pi i \alpha_i)|i\rangle$ with the random state over all Δ*N* sites, random phase $\alpha_i \in [0, 1]$, $|i\rangle \equiv c_i^\dagger |0\rangle$ and summation over all sites of the chosen subsystem Δ*N*. After that, we transform the origin Hamiltonian (1) through redefining it in another (new) basis and set the wave function $|\psi_{rnd}\rangle$ as the first vector in the new basis. Within the tridiagonal representation of the

Hamiltonian and calculating the $\rho_1(E) = -\pi^{-1} \mathrm{Im}\, G_{11}(E+i\zeta)$ *via* the continued fraction technique, we obtain the value $\rho_1(E)$ corresponding to the total DOS per one atom (site) of the lattice at hand. Now, there is no any necessity to calculate the remaining $N-1$ matrix elements $G_{ii}$, thereby we avoid the quadratic dependence of the computational efforts on the system size and keep the scaling linear ($\propto N$).

The size of the phosphorene lattice, which acts in our calculations as a computational domain, includes 1300×1000 atomic sites along the zigzag (*x*) and armchair (*y*) directions, respectively, which corresponds to ≈ 450×450 nm$^2$, *i.e.*, comparable with realistic phosphorene samples treated in experiments. As mentioned above, to study the strain effect, we subject such a lattice to the uniaxial tensile (Fig. 2a, b) or shear (Fig. 2c, d) deformations as well as their combinations (Fig. 2e–h) when both types of strain are applied. Each of them can be oriented along the zigzag (Fig. 2a, c) or armchair (Fig. 2b, d) directions when they are loaded separately (Fig. 2a–d) or simultaneously (Fig. 2e–h). Tension along any other direction can be represented as a combination of zigzag + armchair stretches.

*Ab initio* **computational details**

To evaluate the phosphorene's band structure, — a fundamental information on the electronic properties describing the relationship between the energy and electron wave vectors, — we performed the first principles numerical calculations using the open source computer codes within the Quantum Espresso (QE) electronic-structure simulation package[77]. This package includes the density-functional theory (DFT)[78], plane wave's basis sets and pseudopotentials to represent the electron–ion interactions. The computation package includes[77] the calculation of the Kohn–Sham orbitals and energies, complete structural optimizations of the microscopic (atomic coordinates) and macroscopic (unit cell) degrees of freedom, using Hellmann–Feynman forces and stresses. The projector augmented wave (PAW) technique[79] for generalization of the pseudopotentials, self-consistent total energy calculations and geometry optimization is implemented in the package. The standard generalized gradient approximations (GGA)[80–82] formalism for the Perdew–Burke–Ernzerhof (PBE)[83,84] exchange-correlation functional is adopted. The geometry optimization procedure before proceeding to the calculations was required to reach the atomic structure relaxation state. Calculated lattice parameters after the geometry optimization of unstrained lattice were $|\mathbf{a}_1|$ = 3.344 Å and $|\mathbf{a}_2|$ = 4.588 Å (see Fig. 1b) that is in agreement with previous theoretical predictions[10,11,39,47,48] and the values obtained from experiment[85]. In order to prevent (or at least minimize) interaction between the layers due to the periodic boundary condition, a vacuum layer of 22 Å was included in the unit cell along the transverse direction to the phosphorene layer. Phosphorus atoms have electron configuration $1s^22s^22p^63s^23p^3$, which includes five (valence) electrons at the highest-numbered (third) shell ($3s^23p^3$): two and three of them in the $3s$ and $3p$ subshells, respectively. The electron wave functions were expanded in plane waves with a kinetic energy cut-off of 34.414 Ry, while the energy cut-off for the charge density was set to ten times larger. The Brillouin-zone integration was performed by a 10×8×1 **k**-points grid (mesh) following the scheme proposed by Monkhorst–Pack[86]. To simulate the uniaxial or shear strain, we changed or shifted the lattice parameters and defined the strain magnitude ε through its relation with the relative change of the sides (parameters) as expressed in Figs. 2a–d. The structure was allowed to relax (be geometrically optimized) after each step of the strain magnitude for each type and direction of the deformation.

## Results and discussion

Before proceeding to study of the strained phosphorene sample, in order to test and validate our numerical model, it is reasonably to consider this material initially at ambient condition, *i.e.*, when it is unstrained. The DFT-based *ab initio* calculations (within QE simulation package) of the energy band structure in Fig. 3a show that the single-layer phosphorene is a semiconductor with a nearly direct band gap at the centre of the Brillouin zone (Γ-point). The calculated band gap is circa 0.945 eV, which agrees with the previous first principals computational results based on the GGA/PBE[12,28,30,39,47,48,55]. All electronic bands along the *S–Y* path are double-degenerated. Similar double degeneracy has been also revealed for the phonon dispersion curves (modes)[10,28]. The high dispersions between Γ and *X* are observed in Fig. 3a for both valence and conduction bands. For the $\Gamma$-to-$Y$ direction, the dispersion is smaller in the conduction band, but strong in the valence band, which is changed from a rather flat near the $\Gamma$ point to the deep minimum near the $Y$ point.

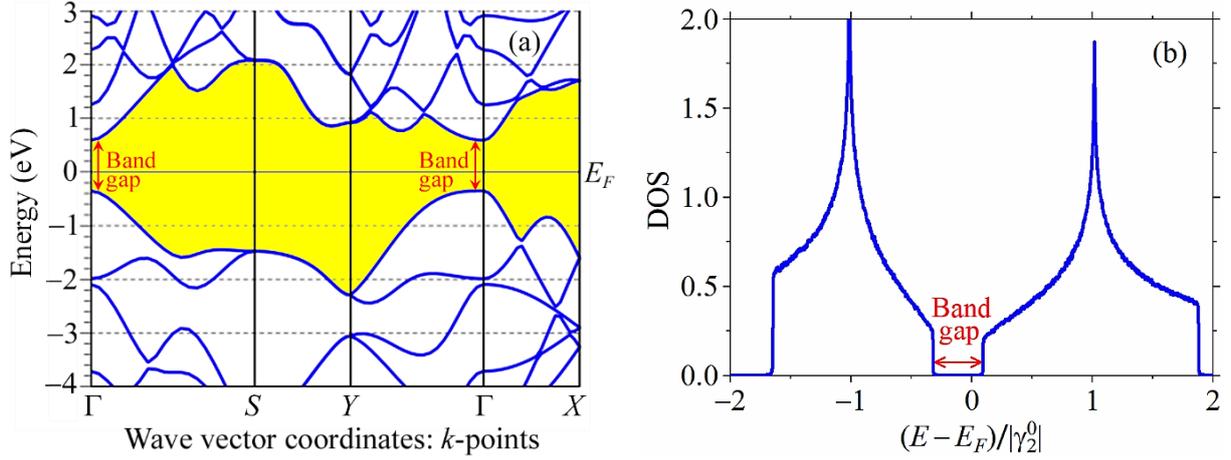

**Figure 3.** The electron energy, $E$, band structure (a) and total density of states (b) of unstrained phosphorene calculated within the standard DFT (GGA-PBE) method and TB model, respectively. Shaded (highlighted in yellow online) region is an area between the valence band maximum (located along the highest curve profile below the Fermi level $E_F$ set to be zero) and conduction band minimum (along the lowest curve profile above the $E_F$).

The band gap value in Fig. 3a contradicts to the gap width, $\approx 1.5$ eV, in Fig. 3b, calculated within the framework of TB model, which in turn practically coincides with the experimentally measured gap value[2] and well agrees with *ab initio* calculations[10,12,27,28,45] based on the of Heyd–Scuseria–Ernzerhof (HSE)06 hybrid functional[87]. Such an agreement justifies the TB Hamiltonian (1) we used in the model and permits of adequate and accurate results that we expect to obtain for the cases of the strain-affected hopping parameters (2). It is well known from the literature that the standard DFT (based on the PBE pseudopotential) calculations underestimate the band gap in phosphorene. This problem can be solved, *e.g.*, through using the hybrid HSE06 functional[10,12,27,28,45] or *GW* (instead of GGA) approximation[48,62,66,67]. Nevertheless, in spite of the gap underestimating, there are no any other significant changes in the electronic structure for the standard global hybrid (PBE) and screened hybrid (HSE06) functionals as well as *GW* correction. Each of them exhibits very similar features and tendencies in the band gap behaviour, including its strain-induced nonmonotonic dependence and direct–indirect gap transition (see, *e.g.*, references in the caption to the figure firstly mentioned in the next paragraph). Therefore, sustaining some other authors[28,47,48], we do not focus on this problem, as that is not the scope of this work.

The band gap evolution in phosphorene subjected to uniaxial tensile and shear strains is presented in Fig. 4, where results of our calculations within the TB model (with distant-dependent hopping parameters) and DFT are compared with other computations, where authors used different pseudopotentials, mainly PBE or HSE. In particular, Figs. 4a, b contain comparative results on the uniaxial tensile strains of Wang *et al.* (2015)[55] and Phuc *et al.* (2018)[47] who used the PBE pseudopotential method; Peng *et al.* (2014)[27], Sa *et al.* (2014)[10], and Li *et al.* (2014)[45] applied the HSE method; and Hernandez *et al.*[48] based on the *GW* approach. Figures 4c, d contain comparative results on the shear strains of Ranawat *et al.* (2015)[30] and Sa *et al.* (2015)[28] based on the PBE pseudopotential, along with results of Sa *et al.* (2015)[28] based of HSE method. Note that except Sa *et al.* (2015)[28] some other authors[10,27,48,45,55] also used both types of the pseudopotentials (standard PBE and hybrid HSE) to compare the output computational results on the tensile deformation. However, we do not reproduce them in Figs. 4a, b in order to avoid the figure overload with excessive details and information.

As Figs. 4a–d demonstrate, the tight-binding model results (black solid squares, ■) lead to a linear strain dependence of the electronic band gap for both types and directions of the deformation within the considered strain percentage (up to 15%). The band gap increases or decreases monotonously as the stretching or compression increases or decreases, respectively (Figs. 4a, b).

These findings agree with the linear strain-induced change of the band gap as a result of the combined (valence-force + tight-binding) multi-scale approach of Midtvedt et al.[63] at small–moderate deformation values (up to 5%) for which the approach is valid. Our TB calculations also sustain numerical results of Midtvedt et al.[63] on the slightly faster linear change of the band gap for armchair stretching (Fig. 4a) in comparison with the zigzag one (Fig. 4b).

In contrast to the TB-model-based results, the DFT-based calculations predict the nonmonotonic behaviour of the band gap as a function of applied uniaxial tensile deformation along either zigzag direction (Fig. 4a) or armchair one (Fig. 4b) independently on the approximation (GGA or GW) and pseudopotential (PBE or HSE06) used. In this respect, our DFT calculations (black solid circles, ●) sustain first principles findings of other authors. We observe the qualitative agreement between our DFT curves and ab initio findings based on HSE pseudopotential[10,27,45] or GW approximation[48] in Figs. 4a, b, and even more or less quantitative coincidence as compared with the PBE-based results[47,55]. The nonmonotonic behaviour of the strain-dependent band gap as well as its direct–indirect transition is attributed to the changes occurring on different regions of the valence band maximum (VBM) and conduction band minimum (CBM) in the band structure (Fig. 3a). As the uniaxial strain increases, the dispersion on the $\Gamma$–$X$ and $\Gamma$–$Y$ modes changes, and new maximums and minimums on these paths of the VBM and CBM, respectively, appear, which results to new real (indirect) band gap size smaller than the direct one (in the $\Gamma$ point)[48]. The complete understanding (explanation) of the reason of the direct–indirect gap transition and its nonmonotone change comes from the studying the contributions to the gap size from the *s*, $p_x$, $p_y$, $p_z$ atomic orbitals on the partial (projected) density of states (PDOS)[48]. With an increase of the strain percentage, the orbitals can mix (overlap) to generate the new hybrid orbitals (*i.e.*, undergo the hybridization) near (below or above) the Fermi level with the formation of the new (valence or conduction) band and gap value[48]. Our additional calculations show that the nonmonotonic band gap behaviour can be realizable within the TB model through changing (increasing in several times) the decay rate parameter β in Eq. (2).

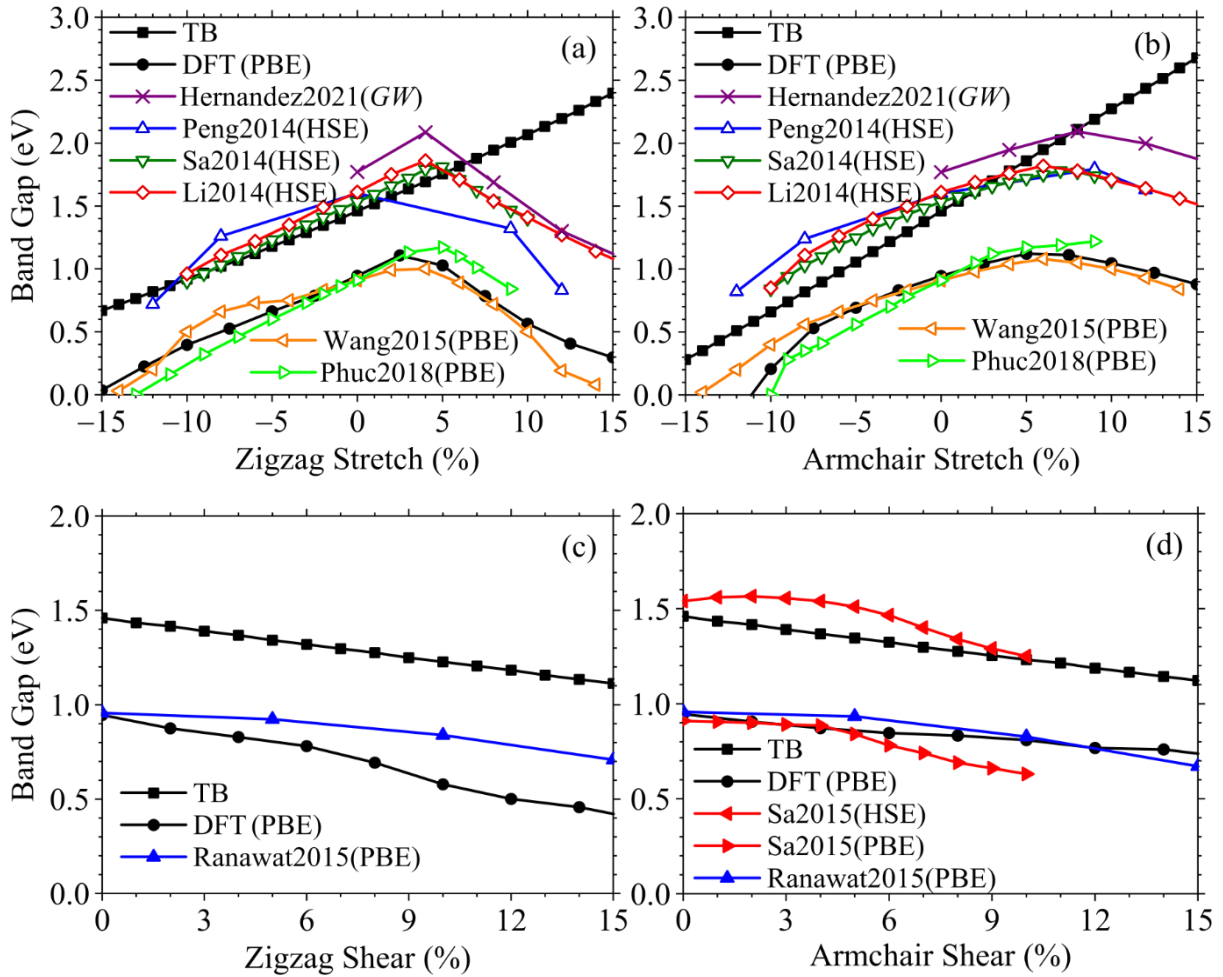

**Figure 4.** Band gap vs. uniaxial tensile (a, b) and shear (c, d) strains calculated within the tight-binding (TB) model (■) and density functional theory (DFT) method (●) as compared with numerical findings of other authors. Here, △—Peng et al. (2014)[27], ▽—Sa et al. (2014)[10], ◇—Li et al. (2014)[45], ◁—Wang et al. (2015)[55], ▷—Phuc et al. (2018)[47], ×—Hernandez et al. (2021)[48], ▲—Ranawat et al. (2015)[30], ◀ and ▶—Sa et al. (2015)[28]. A colour version of this figure is available online.

Figures 4c, d exhibits evidence of the band gap gradual depletion as the shear deformation increases along either zigzag or armchair direction independently on the model and approximation employed. All three approaches (TB, PBE and HSE) employed for the study yield results, which differ from each other only qualitatively, but not quantitatively. It is expected that the curves reflecting the gap values calculated based on the PBE pseudopotential are lower than those ones obtained based on the HSE functional as well as within the framework of the TB model. As in case of

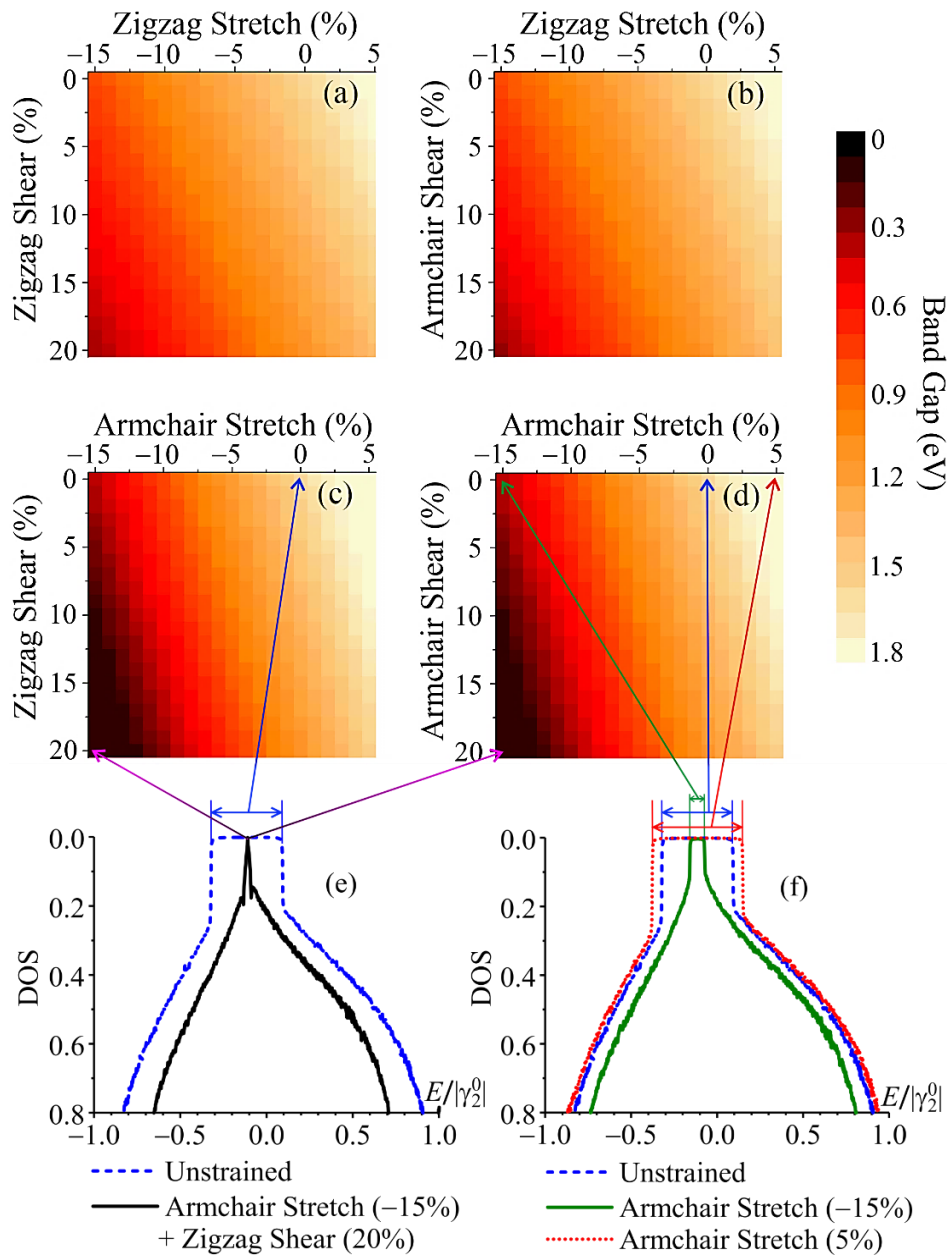

**Figure 5.** (Colour online) Strain-dependent band-gap patterns for black phosphorene under tensile (compression) and shear deformations (a–d). Representative DOS curves (e, f) demonstrate how the diagrams (a–d) are plotted: each band gap value in (a–d) equals to the plateau (gap) width extracted from the corresponding DOS curve as arrows indicate.

the uniaxial strains, the band gap also experiences the direct-to-indirect transition when the certain percentage of the shear stress is loaded, which is revealed based on the DFT calculations[28,30] but cannot be detected within the TB model supplemented with the bond-dependent hopping parameters.

The deformation-dependent band-gap patterns in Figs. 5a–d, exhibiting the effect of simultaneous action of both (uniaxial tensile + shear) strain types, are obtained within the TB model. For this purpose, the family of more than four hundreds of densities of states has been calculated for different values and directions of both deformation types. To plot the diagrams, we extracted the band gap size from the calculated DOS curve as depicted in representative Figs. 5e, f. One can observe from Fig. 5 that the bang gap closes, — the semiconductor-to-(semi)metal phase transition occurs, — when both strain types are applied: ≈15% of the armchair compression + at least ≈10% of shearing along any of two (armchair or zigzag) directions. That is the semiconductor–(semi)metal phase transition can be reached at lower percentages of the strains if both deformation types are switched on. Inasmuch as the phosphorene lattice can be unstable under large deformations (*e.g.*, compression)[10], it is important since offers us a possibility to reach the band gap closing within the range of strictly non-destructive deformations of the phosphorene in case of the necessity of the semiconductor-to-(semi)metal phase transition for the practical application.

## Conclusions

Monolayer black phosphorene, as a still relatively new member of the 2*D* family, is a remarkably suitable object of the study in the field of 2*D* straintronics on the way of following its main concept: engineering the electronic properties of 2*D* materials by means of the introduction of mechanical deformations. Intraplanar strains, namely, uniaxial tension and shearing, are one of them acting as a powerful tool for tuning the electronic band gap from zero up to the values which are inherent to the wide-band-gap semiconductors (up to 2 eV and above), that is much more higher than for silicon (1.12 eV) commonly used in electronics devices.

To study the effects of the uniaxial tensile strain and shear deformation as well as their combinations on the band gap in phosphorene, the tight-binding Hamiltonian with the distant-dependent hopping integrals is used, and obtained results are compared with both our first-principles calculations and those obtained by other authors. TB model allows to perform calculations with the systems containing millions of atoms (as the real experimental samples contain), and thus is substantially less computationally demanding in comparison with any *ab initio* simulations, restricted to periodic supercells or lattice fragments with a relatively small number of atoms due to the computational expensiveness.

The TB-model-based findings indicate that the band gap value of unstrained phosphorene agrees with both the experimental one and that obtained using the DFT with the screened hybrid (HSE06) functional or *GW* correction, and linearly depends on the both deformation types. Namely, the band gap increases (decreases) as the uniaxial tensile strain increases (decreases), and gradually decreases with increasing the shear deformation. As the decay rate parameter defining the bond-length dependence of the hoppings enhances, the linear dependence of the band gap on the uniaxial tensile strain becomes the non-monotonic one and similar to that obtained from our and other authors' DFT calculations. The nonmonotonic dependence of the band gap with the stretching deformation can be understood from the partial DOS exhibiting the contributions to the gap size from different ($s$, $p_x$, $p_y$, $p_z$) atomic orbitals that can overlap and hybridize near the Fermi level with the formation of the new band and gap size as the strain percentage growths. Within the TB model, we extract the information on the band gap width from the total DOS, therefore cannot detect the overlapping or hybridization. In case of the shear strain, both (TB and DFT) methods give the gradual degradation of the gap with only difference in its value, which is underestimated for DFT-PBE as compared to the DFT-HSE and TB.

The strain-dependent band-gap pattern diagrams demonstrate the variety of the continuous gap values realizable when a combined (tensile/compression + shear) strain loaded. A semiconductor-to-semimetal phase transition in the phosphorene can be reached at lower percentages of the strains if both deformation types (stretching + shearing) are switched on. It offers us a possibility to reach the bandgap closing at lower (strictly non-destructive) deformations of the phosphorene, if such a transition appears to be useful for overcoming of the challenges dealing with modification of its properties and functionalization.

## Competing interests

The authors declare no competing interests.

## Acknowledgements

The first author acknowledges the National Academy of Sciences of Ukraine for support within the framework of the program of post-doctoral research in the N.A.S. of Ukraine for 2021–2023 through the project "Complex diagnostics of sensitive to strains and defects structural and electronic properties of metallic nanomaterials" (state reg. No. 0120U102265). The third and fourth authors acknowledge the National Academy of Sciences of Ukraine for support within the departmental research for 2022–2026 (state reg. No. 0122U002396). All authors obliged to the Armed Forces of Ukraine for providing security made possible to perform this work.

## Author contributions

A.G.S. and I.Y.S. performed numerical calculations using the DFT-based QE simulation package and TB-model-based computations, respectively. T.M.R. designed the project, reviewed the literature, collected data, supervised the findings of this work, and wrote the manuscript with input from all authors. V.A.T. supervised the project, devised the main conceptual ideas, verified analytical approaches, and provided critical feedback. All authors were in charge of the overall direction and planning, analysed and discussed the results, commented on the manuscript, and contributed to its final version.

## Data availability

The data that support the findings of this study available from the corresponding author on reasonable request.